\documentstyle[12pt]{article}

\catcode`\@=11

\@addtoreset{equation}{section}
\def\theequation{\thesection\arabic{equation}}

\def\@normalsize{\@setsize\normalsize{15pt}\xiipt\@xiipt
\abovedisplayskip 14pt plus3pt minus3pt%
\belowdisplayskip \abovedisplayskip
\abovedisplayshortskip  \z@ plus3pt%
\belowdisplayshortskip  7pt plus3.5pt minus0pt}
\def\small{\@setsize\small{13.6pt}\xipt\@xipt
\abovedisplayskip 13pt plus3pt minus3pt%
\belowdisplayskip \abovedisplayskip
\abovedisplayshortskip  \z@ plus3pt%
\belowdisplayshortskip  7pt plus3.5pt minus0pt
\def\@listi{\parsep 4.5pt plus 2pt minus 1pt
            \itemsep \parsep
            \topsep 9pt plus 3pt minus 3pt}}

\def\underline#1{\relax\ifmmode\@@underline#1\else
        $\@@underline{\hbox{#1}}$\relax\fi}
\@twosidetrue
\relax

\catcode`@=12

\catcode`\@=11

\def\section{\@startsection{section}{1}{\z@}{3.5ex plus 1ex minus
   .2ex}{2.3ex plus .2ex}{\large\bf}}
\def\thesection{\arabic{section}.}

\def\ps@headings{\def\@oddfoot{}\def\@evenfoot{}
\def\@oddhead{\hbox{}\hfill
        \makebox[.5\textwidth]{\raggedright\ignorespaces --\thepage{}--
        \hfill }}
\def\@evenhead{\@oddhead}
\def\subsectionmark##1{\markboth{##1}{}}
}

\ps@headings

\catcode`\@=12

\relax

\def\figcap{\section*{Figure Captions\markboth
        {FIGURECAPTIONS}{FIGURECAPTIONS}}\list
        {Fig. \arabic{enumi}:\hfill}{\settowidth\labelwidth{Fig. 999:}
        \leftmargin\labelwidth
        \advance\leftmargin\labelsep\usecounter{enumi}}}
 \relax
\def\tablecap{\section*{Table Captions\markboth
        {TABLECAPTIONS}{TABLECAPTIONS}}\list
        {Table \arabic{enumi}:\hfill}{\settowidth\labelwidth{Table 999:}
        \leftmargin\labelwidth
        \advance\leftmargin\labelsep\usecounter{enumi}}}
 \relax
\def\reflist{\section*{References\markboth
        {REFLIST}{REFLIST}}\list
        {[\arabic{enumi}]\hfill}{\settowidth\labelwidth{[999]}
        \leftmargin\labelwidth
        \advance\leftmargin\labelsep\usecounter{enumi}}}
 \relax

\catcode`\@=11

\def\marginnote#1{}
\newcount\hour
\newcount\minute
\newtoks\amorpm
\hour=\time\divide\hour by60
\minute=\time{\multiply\hour by60 \global\advance\minute by-
\hour}
\edef\standardtime{{\ifnum\hour<12 \global\amorpm={am}%
    \else\global\amorpm={pm}\advance\hour by-12 \fi
    \ifnum\hour=0 \hour=12 \fi
    \number\hour:\ifnum\minute<100\fi\number\minute\the\amorpm}}
\edef\militarytime{\number\hour:\ifnum\minute<100\fi\number\minute}
\def\draftlabel#1{{\@bsphack\if@filesw {\let\thepage\relax
  \xdef\@gtempa{\write\@auxout{\string
    \newlabel{#1}{{\@currentlabel}{\thepage}}}}}\@gtempa
    \if@nobreak \ifvmode\nobreak\fi\fi\fi\@esphack}
     \gdef\@eqnlabel{#1}}
\def\@eqnlabel{}
\def\@vacuum{}
\def\draftmarginnote#1{\marginpar{\raggedright\scriptsize\tt#1}}
\def\draft{\oddsidemargin -.5truein
        \def\@oddfoot{\sl preliminary draft \hfil
        \rm\thepage\hfil\sl\today\quad\militarytime}
        \let\@evenfoot\@oddfoot \overfullrule 3pt
        \let\label=\draftlabel
        \let\marginnote=\draftmarginnote
   
\def\@eqnnum{(\theequation)\rlap{\kern\marginparsep\tt\@eqnlabel}%
\global\let\@eqnlabel\@vacuum}  }
\def\preprint{\twocolumn\sloppy\flushbottom\parindent 1em
        \leftmargini 2em\leftmarginv .5em\leftmarginvi .5em
        \oddsidemargin -.5in    \evensidemargin -.5in
        \columnsep 15mm \footheight 0pt
        \textwidth 250mmin      \topmargin  -.4in
        \headheight 12pt \topskip .4in
        \textheight 175mm
        \footskip 0pt
        
\def\@oddhead{\thepage\hfil\addtocounter{page}{1}\thepage}
        \let\@evenhead\@oddhead \def\@oddfoot{} \def\@evenfoot{} 
}
\def\titlepage{\@restonecolfalse\if@twocolumn\@restonecoltrue\onecolumn
     \else \newpage \fi \thispagestyle{empty}\c@page\z@
        \def\thefootnote{\fnsymbol{footnote}} }
\def\endtitlepage{\if@restonecol\twocolumn \else  \fi
        \def\thefootnote{\arabic{footnote}}
        \setcounter{footnote}{0}}  
\catcode`@=12
\relax

\def\ps@headings{\def\@oddfoot{}\def\@evenfoot{}
\def\@oddhead{\hbox{}\hfill
        \makebox[.5\textwidth]{\raggedright\ignorespaces --\thepage{}--
        \hfill }}
\def\@evenhead{\@oddhead}
\def\subsectionmark##1{\markboth{##1}{}}
}

\ps@headings

\relax

\def\firstpage#1#2#3#4#5#6{
\begin{document}
\begin{titlepage}
\nopagebreak
\title{\begin{flushright}
        \vspace*{-1.8in}
        {\normalsize CERN--TH/97-43}\\[-9mm]
       {\normalsize NUB--#1 #2}\\[-9mm]
   {\normalsize CPTH--S498.0397}\\[-9mm]
        {\normalsize hep-th/9703076}\\[4mm]
\end{flushright}
\vfill
{#3}}
\author{\large #4 \\[1.0cm] #5}
\maketitle
\vskip -7mm     
\nopagebreak 
\begin{abstract}
{\noindent #6}
\end{abstract}
\vfill
\begin{flushleft}
\rule{16.1cm}{0.2mm}\\[-3mm]
$^{\star}${\small Research supported in part by\vspace{-4mm}
the National Science Foundation under grant
PHY--96--02074, \linebreak and in part by the EEC \vspace{-4mm} 
under the TMR contract    
ERBFMRX-CT96-0090 and under CHRX-CT93-034.}\\
$^{\dagger}${\small Laboratoire Propre du CNRS UPR A.0014.}\\
March 1997
\end{flushleft}
\thispagestyle{empty}
\end{titlepage}}

\def\simlt{\stackrel{<}{{}_\sim}}
\def\simgt{\stackrel{>}{{}_\sim}}
\newcommand{\dal}{\raisebox{0.085cm}
{\fbox{\rule{0cm}{0.07cm}\,}}}
\newcommand{\dt}{\partial_{\langle T\rangle}}
\newcommand{\dtbar}{\partial_{\langle\bar{T}\rangle}}
\newcommand{\al}{\alpha^{\prime}}
\newcommand{\mst}{M_{\scriptscriptstyle \!S}}
\newcommand{\mpl}{M_{\scriptscriptstyle \!P}}
\newcommand{\dv}{\int{\rm d}^4x\sqrt{g}}
\newcommand{\lv}{\left\langle}
\newcommand{\rv}{\right\rangle}
\newcommand{\ph}{\varphi}
\newcommand{\abar}{\bar{a}}
\newcommand{\sbar}{\,\bar{\! S}}
\newcommand{\xbar}{\,\bar{\! X}}
\newcommand{\fbar}{\,\bar{\! F}}
\newcommand{\zbar}{\bar{z}}
\newcommand{\dbar}{\,\bar{\!\partial}}
\newcommand{\tbar}{\bar{T}}
\newcommand{\taubar}{\bar{\tau}}
\newcommand{\ubar}{\bar{U}}
\newcommand{\tetabar}{\bar\Theta}
\newcommand{\etabar}{\bar\eta}
\newcommand{\qbar}{\bar q}
\newcommand{\ybar}{\bar{Y}}
\newcommand{\phb}{\bar{\varphi}}
\newcommand{\cm}{Commun.\ Math.\ Phys.~}
\newcommand{\prl}{Phys.\ Rev.\ Lett.~}
\newcommand{\pr}{Phys.\ Rev.\ D~}
\newcommand{\pl}{Phys.\ Lett.\ B~}
\newcommand{\ibar}{\bar{\imath}}
\newcommand{\jbar}{\bar{\jmath}}
\newcommand{\np}{Nucl.\ Phys.\ B~}
\newcommand{\F}{{\cal F}}
\renewcommand{\L}{{\cal L}}
\newcommand{\A}{{\cal A}}
\newcommand{\M}{{\cal M}}
\newcommand{\K}{{\cal K}}
\newcommand{\T}{{\cal T}}
\renewcommand{\Im}{\mbox{Im}}
\newcommand{\e}{{\rm e}}
\newcommand{\be}{\begin{equation}}
\newcommand{\en}{\end{equation}}
\newcommand{\gsi}{\,\raisebox{-0.13cm}{$\stackrel{\textstyle
>}{\textstyle\sim}$}\,}
\newcommand{\lsi}{\,\raisebox{-0.13cm}{$\stackrel{\textstyle
<}{\textstyle\sim}$}\,}
\date{}
\firstpage{3155}{}
{\large\bf Duality of N{=}2 Heterotic -- Type I Compactifications\\[-3mm] 
in Four Dimensions$^{\star}$} 
{I. Antoniadis$^{\,a,b}$, H. Partouche$^{\,a}$
and T.R. Taylor$^{\,c}$}  
{\normalsize\sl
$^a$Centre de Physique Th{\'e}orique, Ecole Polytechnique,$^\dagger$
{}F-91128 Palaiseau, France\\[-3mm]
\normalsize\sl $^b$Theory Division, CERN, 1211 Geneva 23,
Switzerland\\[-3mm]
\normalsize\sl $^c$Department of Physics, Northeastern
University, Boston, MA 02115, U.S.A.}
{We discuss type I -- heterotic duality in four-dimensional
models obtained as a Coulomb
phase of the six-dimensional $U(16)$ orientifold
model compactified on $T^2$ with arbitrary $SU(16)$ Wilson lines. 
We show that K{\"a}hler potentials, gauge threshold
corrections and the infinite tower of higher derivative F-terms  
agree in the limit
that corresponds to weak coupling, large $T^2$ heterotic
compactifications. On the type I side,
all these quantities are completely determined by the spectrum
of $N{=}2$ BPS states that originate from $D{=}6$ massless superstring modes.}
\section{Introduction}

In a previous paper \cite{ABFPT}
we presented a general computation of the perturbative
prepotential in four-dimensional type I string vacua with $N=2$ space-time
supersymmetry. As we have shown, its one-loop contribution is determined
entirely by the corresponding correction to the Planck mass. Finally, we
applied the result to study duality in the so-called $STU$-model with
three 
vector multiplets, 
which admits simultaneous type II and heterotic descriptions. This model,
on the type I and heterotic side, corresponds to the Higgs phase of a
six-dimensional vacuum with gauge group $U(16)$ \cite{gp,bl}, 
compactified to four dimensions on $T^2$.

In this work we study heterotic -- type I duality \cite{pw}
in the context of the original $U(16)\times U(1)^3$ model. More precisely,
we consider the Coulomb phase of the above model whose massless
spectrum consists of $n_V=19$ Abelian vector multiplets and $n_H=4$
neutral
hypermultiplets. Two vector multiplets, $S$ and $S'$,  play a special
role in type I theory: their scalar vacuum expectation values
determine the tree-level gauge couplings of the vector bosons associated
to $U(16)$. In this class of models, the perturbative 
prepotential has the form \cite{ABFPT}:
\be
{\F}^I=SS'U-\frac{1}{2}\sum_i(v_iS+v'_iS')A_i^2+f^I(U,A)\ , \label{prei}
\en
where $v_i$ and $v'_i$ are constants and $f^I(U,A)$ is the 
one-loop correction. Type I -- heterotic
duality maps $S$, $S'$ and $U$ into the ``universal'' $S$, $T$ and $U$
heterotic multiplets, respectively.

On the heterotic side, the perturbative prepotential has the form
\be
{\F}^H=STU-\frac{1}{2}\sum_iv_i\, SA_i^2+f^H(T,U,A)\ ,  \label{preh}
\en
where the constants $v_i$ are given by the
Ka\v{c}-Moody levels of the corresponding gauge groups and 
$f^H(T,U,A)$ is the one-loop correction. The limits Im$T\to\infty$
and the respective Im$S>~$Im$S'\to\infty$ take both theories into 
perturbative regimes, where it is possible to test duality by comparing 
perturbative prepotentials and/or the related
K{\"a}hler metrics and gauge threshold corrections \cite{ABFPT}. 
In particular, on the heterotic side,
such a limit corresponds to a weakly coupled theory compactified on
large $T^2$. For a dual pair of models
\be
f^H(T,U,A)~\stackrel{T\to\infty}{\longrightarrow}-{1\over 2}\sum_iv'_i\, TA_i^2
+f^I(U,A)\ . \label{pretest}
\en
Note that a part of the one-loop heterotic prepotential is mapped
into the tree-level type I couplings proportional to $v'_i$
while the remaining part corresponds to type I one-loop corrections.
This is of course also true for the related K{\"a}hler metrics
and gauge threshold corrections, which can be extracted
directly from the appropriate superstring scattering amplitudes.
In the models under consideration, $v'_i=0$ for all gauge fields
associated to the $SU(16)$ subgroup of $U(16)$.

Prepotential is only the first representative of a large class
of analytic quantities characterizing $N{=}2$ supersymmetric effective
field
theories. Effective Lagrangians contain also 
higher derivative F-terms of the form $I_g={\F}_gW^{2g}$,
where $W$ is the Weyl superfield. 
The prepotential corresponds to $\F_0$ while higher $\F_g$'s
correspond to analytic couplings associated to higher derivative
interactions
\cite{top}.
In ref.~\cite{AGNT}, type II -- heterotic duality was tested by comparing
$\F_g$'s of dual models,
while in ref.~\cite{Se} a similar analysis was done to test 
type I -- heterotic duality in the $STU$-model. 
In this work, we apply similar tests to type I -- heterotic duality,
in the Coulomb phase of the $U(16)\times U(1)^3$ model 
{\em i.e}.\ in the presence of non-vanishing Wilson lines.
On the type II side, $\F_g$'s are subject to an exact non-renormalization 
theorem, which implies that they are purely genus-$g$ quantities.
On the other hand, for type I and heterotic, all $\F_g$'s with $g\ge 2$
start appearing at the one-loop level. 
Duality can then be tested by comparing
them in the weak coupling limits discussed before.
An important difference between the type II -- heterotic and type I --
heterotic
comparisons is that in the first case the tests are restricted to the
holomorphic anomaly equation and to the behaviour of $\F_g$'s near singular
points of the moduli space (such as conifolds), while in the latter the
comparison is extended to their full expression in the appropriate limits.

The case of $\F_1$, which is related to the gravitational $R^2$
threshold corrections \cite{AGN}, is a little more subtle. 
Due to non-localities of the effective action produced
by one-loop anomalies, $\F_1$ cannot be defined in a straightforward
manner; this is exactly the same problem as was first encountered in
the analysis of $N=1$ threshold corrections \cite{dkl} where it was shown
that, in a strict sense, the gauge function does not exist.
In this paper, we adopt the same convention as in ref.~\cite{AGNT},
with $\F_1$ defined as the coefficient of the $R^2$ term.
For a heterotic model,
\be
\F_1^H=4\pi\makebox{Im}S+f_1^H(T,U,A),   \label{gravh}
\en
while for a type I model
\be
\F_1^I=4\pi\makebox{Im}S+4\pi v'_{\mbox{\scriptsize grav}}
\makebox{Im}S'+f_1^I(U,A)\ ,
\label{gravi}
\en
where $v'_{\mbox{\scriptsize grav}}$ is a constant.
For a dual pair,
\be
f_1^H(T,U,A)~\stackrel{T\to\infty}{\longrightarrow}~4\pi 
v'_{\mbox{\scriptsize grav}}\makebox{Im}T +f_1^I(U,A)\ ,
\label{gravtest}
\en
so similarly to the prepotential, a part of one-loop
heterotic corrections is mapped into the tree-level
type I coupling. In the model under consideration,
$v'_{\mbox{\scriptsize grav}}=1$ consistently with $T$-duality
of the type I vacuum, which inverts the volume of $K_3$ \cite{ABFPT}.

The paper is organized as follows.
In section 2, we determine the moduli K{\"a}hler metrics for 
the $U(16)\times U(1)^3$ type I model with $SU(16)$ Wilson lines.
In section 3 we give the free-fermionic heterotic description of this
model and compute its K{\"a}hler metrics.
We show that the heterotic and type I results agree
in appropriate limits. We also make a comparison
of the $SU(16)$ gauge group threshold corrections, which were computed on
the
type I side in ref.~\cite{BF}.
In section 4 we discuss $\F_g$ couplings,
showing that the whole infinite towers of higher
derivative interactions are identical for the dual pairs of models;
this provides an overwhelming evidence for type I -- heterotic duality.

\section{Type I model}

The model we consider on the type I side is constructed as an orientifold
of type IIB \cite{gp} compactified 
on $T^4/{\bf Z}_2\times T^2$. Its massless spectrum
from closed strings consists of the $N=2$ gravity multiplet, 
three Abelian vector
multiplets and four neutral hypermultiplets in the untwisted sector,
as well as 16 additional singlet twisted hypermultiplets. On the
other hand, one has open strings with ends moving on $2\times 16$
nine-branes
and/or $2\times 16$ five-branes, giving rise to a $U(16)$ gauge group
together
with two hypermultiplets in the ${\bf 120}_{1/2}$ representation 
(from the 99-sector)
and sixteen in the ${\bf 16}_{1/4}$ of $U(16)$ (from the 95-sector).
Moreover,
in the case when the five-branes are located at different fixed points of
the orbifold, there is a $U(1)^{16}$ gauge group from the 55-sector (which
eventually become massive by coupling to the sixteen twisted
hypermultiplets).

In the following, we consider the Coulomb phase of the 99-gauge group by
turning on the sixteen Wilson lines on $T^2$. One thus obtains 3+16
Abelian 
vector multiplets whose moduli space in $N=2$ special coordinates is
parametrized by the complex fields $S$, $S'$, $U$ and
$A_i=a_4^i-a_5^iU$ \cite{ABFPT}.
As usual, $U$ parametrizes the complex structure of $T^2$ while
$a^i_{4,5}$
denote the sixteen (real) Wilson lines along the two compact directions 
(4 and 5) of $T^2$. The fields
$S$ and $S'$ are appropriate combinations involving
the dilaton and the volumes of $T^2$ and $T^4$. Their real parts are
associated
to perturbative continuous Peccei-Quinn (PQ) symmetries 
while their imaginary parts
contain a factor of the inverse string coupling and go to infinity in the
weak
coupling. Under type I -- heterotic duality, they are mapped into the
heterotic
dilaton and K{\"a}hler modulus of the $T^2$, $S$ and $T$ respectively.

Due to analyticity and PQ symmetries, the perturbative prepotential
receives 
one-loop corrections only, that depend on $U$ and $A_i$. Using standard
$N=2$ 
formulae, these can be extracted from the one-loop K{\"a}hler metric
$K^{I(1)}_{U\ubar}$, which in type I strings is determined
uniquely by the one-loop corrections to the Einstein kinetic term.
Applying the general results of ref.~\cite{ABFPT}, we find\footnote{Here,
the
Wilson lines correspond to the $SU(16)$ generators and are subject to the 
condition $\sum_ia^i_{4,5}=0$. The $U(1)$
  factor, which is anomalous in six dimensions, requires special
  treatment \cite{d,bl} and will not be discussed in the present work.}
\be
K^{I(1)}_{U\ubar}={{\sqrt G}\over 16\pi^2
\mbox{Im}S'}\partial_U\partial_{\ubar}
\int_0^\infty {dt\over t^2}\left\{\sum_{{a^I+a^J+\Gamma_2}}s_{IJ}
+\sum_{{2a^I+\Gamma_2}} -16\sum_{{a^I+\Gamma_2}} \right\}
e^{-\pi t|p|^2/2{\sqrt G}}\ ,
\label{KI}
\en
where ${\sqrt G}$ is the volume of $T^2$, and $\Gamma_2$ is the
two-dimensional 
lattice of Kaluza-Klein momenta $G^{-1/4}p$, with
\be
p={m_4-m_5 U\over \sqrt{2\mbox{Im}U}}\ ,
\label{mom}
\en
and integer $m_4$, $m_5$. The lattice is shifted by the Wilson lines
as indicated in eq.~(\ref{KI}): for instance, the shift
$a^I+\Gamma_2$ 
implies that 
$m_{4,5}\to m_{4,5}+a^I_{4,5}$. For convenience, we have introduced the
index $I\equiv i$ or $\ibar$, with $i$ and $\ibar$ running over the ${\bf
16}$ 
and ${\bf \overline{16}}$ of $SU(16)$, respectively, and $a^{\ibar}\equiv
-a^i$.
Also, $s_{IJ}=-1$ or 1, depending on whether $I$ and $J$ belong to
the same or conjugate representations. 
Finally, the partial derivatives with respect to $U$ and $\ubar$ are taken
by
keeping the Wilson lines $a_{4,5}^I$ fixed. Note that the volume ${\sqrt
G}$ 
of $T^2$ drops out after a rescaling of $t$.

The three terms in the r.h.s. of 
eq.~(\ref{KI}) correspond to the contributions of the annulus in the
99-sector,
M{\"o}bius strip in the 99-sector and annulus in the 95-sector,
respectively,
while the numerical coefficient of the latter comes from the multiplicity
of
5-branes. The torus contribution vanishes because of extended supersymmetry,
whereas the Klein bottle and annulus in the 55-sector contributions cancel
one another. Note the similarity of the above expression with the one
giving the threshold corrections to gauge couplings \cite{BF}, where
the eigenvalues of the charge-squared operator $(q_I+q_J)^2$, 
$4q_I^2$ and $q_I^2$ are inserted in the three terms of eq.~(\ref{KI}) 
correspondingly.

Using the identity
\be
\partial_U\partial_{\ubar}e^{-\pi t|p|^2/2}
=-\frac{1}{(U-\ubar)^2}t\partial_t^2 te^{-\pi t|p|^2/2}\ ,
\label{id}
\en
which follows from eq.~(\ref{mom}), and performing an integration by
parts, we
find that the boundary term vanishes and
\be
K^{I(1)}_{U\ubar}=-{1\over(U-\ubar )^2}
{1\over 16\pi^2 \mbox{Im}S'}
\int_0^\infty {dt\over t^2}\partial_t 
\left\{ \sum_{a^I+a^J+\Gamma_2}
\!\!\!\!\!s_{IJ}\;
+\sum_{2a^I+\Gamma_2} -\; 16\sum_{a^I+\Gamma_2}
\right\}t\ e^{-\pi t|p|^2/2}\ .
\label{KI1}
\en

\section{Heterotic realization and threshold corrections}

The above model also has a perturbative heterotic description as $SO(32)$
or $E_8\times E_8$ compactified on $T^4/{\bf Z}_2\times T^2$ with
instanton 
numbers 24 or $(12,12)$, respectively. Note that the $T^4$ moduli
belong to neutral hypermultiplets, so they do not appear in the
K{\"a}hler metric of vector moduli. Hence, without loss of generality,
we can consider the fermionic point that corresponds to particular
values of $T^4$ radii. In six dimensions, the model is then
generated by three vectors of boundary
conditions for the world-sheet fermions: the identity ${\bf 1}$, the
supersymmetry vector ${\bf S}$ and a vector ${\bf b}$ that breaks half
of the supersymmetries, as well as the gauge group $SO(32)$ down to
$U(16)$.
However, at the fermionic point, one also obtains an additional $SO(4)^2$
gauge group factor. 

Following refs.~\cite{ABK}, we denote by
$(\partial X^{\mu},\psi^{\mu})$ the left-moving supercoordinates and
$y^{6,...,9}$, $\omega^{6,...,9}$, $\chi^{6,...,9}$ the remaining
left-moving real fermions. The right-movers are $\bar{\partial} X^{\mu}$
together with the real fermions $\bar{y}^{6,...,9},\bar{\omega}^{6,...,9}$
and the complex fermions
$\bar{\eta}^i$, $(i=1,...,16)$. In this notation,
\begin{eqnarray}
{\bf S} &=& \{\psi^\mu, \chi^{6,...,9}\}\nonumber\\
{\bf b} &=& \{\chi^{6,...,9}, y^{6,...,9}; {\bar y}^{6,...,9},
\underbrace{{1\over 2},...,{1\over 2}}_{\bar\eta^i}\}\ ,
\label{basis}
\end{eqnarray}
where the fermions present in the sets are periodic with the exception of 
$\bar{\eta}^i\rightarrow -i\bar{\eta}^i$, while the remaining are
antiperiodic 
under parallel transport along the string. Note that the vectors ${\bf
  1}$, ${\bf S}$ and $2{\bf b}$ generate the untwisted sector of the
$T^4/{\bf Z}_2$ orbifold at the fermionic point, giving rise to
$SO(32)\times SO(8)$ with $N=2$ supersymmetry in $D=6$. 
The matter multiplets in the massless spectrum consist of two 
${\bf 120}_{1/2}$'s
of $U(16)$ from the vectors ${\bf 0}$, ${\bf S}$ of the untwisted
orbifold sector, sixteen ${\bf 16}_{1/4}$'s 
from the vectors $\pm{\bf b}$, ${\bf S}\pm{\bf b}$, 
${\bf 1}+{\bf S}\pm{\bf b}$, ${\bf 1}\pm{\bf b}$ of the twisted
orbifold sector, as well as an additional
untwisted hypermultiplet in the ({\bf 4,4}) representation of $SO(4)^2$.
The
latter can be higgsed away, giving rise to four singlet hypermultiplets,
while
$SU(16)$ can be broken to $U(1)^{15}$ by turning on the Wilson lines upon
toroidal compactification to four dimensions.

The one-loop corrections to the K{\"a}hler metric take the general form
\cite{yuka}:
\be
K^{H(1)}_{U\ubar}={i\over
32\pi^2(U-\ubar)^2}\sum_s\sum_{P\in\Gamma^{(2,18)}_s}
\int_{\cal F}{d^2\tau\over\tau_2^2}{\bar F}_P^{(s)}(\taubar )
\partial_{\taubar}\left( \tau_2
e^{i\pi\tau|p_L|^2}e^{-i\pi\taubar|p_R|^2}\right)\ ,
\label{KH}
\en
where $\tau=\tau_1+i\tau_2$ is the modular parameter of the world-sheet
torus
which is integrated inside its fundamental domain ${\cal F}$, and the sum
is extended over the momenta $P$ of three Lorenzian lattices
$\Gamma^{(2,18)}_s$ associated to the three possible
${\bf Z}_2$ orbifold boundary conditions, $s\equiv (1,0), (0,1), (0,0)$ 
(1 and 0 stand correspondingly for periodic and antiperiodic
coordinates, or equivalently for fermion bilinears
$y\omega\equiv\partial X$).
${\bar F}_P^{(s)}(\taubar )$ is the contribution to the
partition function of the right-moving (bosonic) sector of the heterotic 
superstring, with the exception of the $T^2$ zero modes, which are the left 
and right momenta $p_L,p_R$. Note that the untwisted $N=4$ sector
$s=(1,1)$ gives  vanishing contribution, so that
the result is independent of the $T^4$ moduli as mentioned before.

For vanishing Wilson-lines, the right-moving partition
functions are factorized
\be
\sum_{P\in\Gamma^{(2,18)}_s}\left. {\bar F}_P^{(s)}(\taubar )
e^{i\pi\tau|p_L|^2}e^{-i\pi\taubar|p_R|^2} \right|_{a^i_{4,5}=0}=
{\bar F}^{(s)}Z^{(2,2)}\, ,
\label{F0}
\en
where
\be
Z^{(2,2)}=
\sum_{p_L,p_R\in\Gamma^{(2,2)}}e^{i\pi\tau|p_L|^2}e^{-i\pi\taubar|p_R|^2}\
,
\label{Z22}
\en
and
\begin{eqnarray}
p_L&=&{1\over{\sqrt{2\mbox{Im}T\mbox{Im}U}}}(m_4-m_5U-n_5T-n_4TU)\nonumber\\
p_R&=&{1\over{\sqrt{2\mbox{Im}T\mbox{Im}U}}}(m_4-m_5\ubar-n_5T-n_4T\ubar)\
.
\label{mom22}
\end{eqnarray}
${\bar F}^{(s)}$ can be obtained in a straightforward way by using the
defining basis vectors (\ref{basis}):
\begin{eqnarray}
{\bar F}^{(1,0)}&=&\ {1\over\etabar^4}{4\etabar^2\over
\tetabar^2{1\atopwithdelims()0}}
{\tetabar^{16}{0\atopwithdelims()1/2}+
\tetabar^{16}{0\atopwithdelims()-1/2}-
\tetabar^{16}{1\atopwithdelims()1/2}-
\tetabar^{16}{1\atopwithdelims()-1/2}
\over\etabar^{16}}\nonumber\\
{\bar F}^{(0,1)}&=&-{1\over\etabar^4}{4\etabar^2\over
\tetabar^2{0\atopwithdelims()1}}
{\tetabar^{16}{1/2\atopwithdelims()0}-
\tetabar^{16}{1/2\atopwithdelims()1}+
\tetabar^{16}{-1/2\atopwithdelims()0}-
\tetabar^{16}{-1/2\atopwithdelims()1}
\over\etabar^{16}}\label{Fbar}\\
{\bar F}^{(0,0)}&=&i{1\over\etabar^4}{4\etabar^2\over
\tetabar^2{0\atopwithdelims()0}}
{\tetabar^{16}{1/2\atopwithdelims()1/2}-
\tetabar^{16}{1/2\atopwithdelims()-1/2}-
\tetabar^{16}{-1/2\atopwithdelims()1/2}+
\tetabar^{16}{-1/2\atopwithdelims()-1/2}
\over\etabar^{16}}\ ,\nonumber
\end{eqnarray}
where 
\be
\Theta{\varepsilon\atopwithdelims()\varepsilon'} = \sum_{k\in{\bf Z}} 
e^{i\pi (k+{\varepsilon\over 2})^2\tau+i\pi k\varepsilon'}\ ,
\label{theta}
\en
and $\eta$ is the Dedekind eta function. The three factors in the r.h.s.
of
each ${\bar F}^{(s)}$ in eq.~(\ref{Fbar}) correspond to the
contribution of the (right-moving) space-time and $T^2$ oscillator
modes, the $T^4$ twisted coordinates, and the sixteen complex fermions
${\bar\eta}^i$ generating the $U(16)$, respectively. 

The Wilson lines can now be introduced as boosting parameters of the three
(2,2+16) Lorentzian lattices, in a way that when they vanish we recover
the
above factorized form (\ref{F0}):
$\Gamma_s^{(2,18)}|_{a^i=0}=\Gamma_s^{(16)}\oplus\Gamma^{(2,2)}$. 
This can be implemented by going from 6 to 4 dimensions using
coordinate dependent
compactification, which amounts to changing the boundary conditions
in a way that respects modular invariance \cite{FKPR}. 
In ref.~\cite{FKPR}, this procedure was presented for a compactification 
on $S^1$. It is not difficult to extend these results in the case of $T^2$ 
and find that in the presence of the 16 Wilson lines $a^i_{4,5}$, 
each term of eq.~(\ref{F0}) that is of the form 
$Z^{(2,2)} \bar{\Theta}{u\atopwithdelims()v}^{16}$ (for $u,v=0,1,\pm 1/2$)
is
modified:
\begin{eqnarray}
Z^{(2,2)} \bar{\Theta}{u\atopwithdelims()v}^{16}
\rightarrow{\mbox{Im}T\over\tau_2}\sum_{A=\left( {-n_5\atop -n_4}{l_5\atop
l_4}\right) } e^{2i\pi\bar{T} \det A} e^{-{\pi\mbox{\tiny Im}T
\over \tau_2\mbox{\tiny Im}U}\left| (1,U)A
{\tau\atopwithdelims()1}\right|^2} &\times& \label{modif} \\
\times~  \prod_{i=1}^{16} e^{-i \pi u (l_4 a^i_4+l_5 a^i_5)}e^{i{\pi \over
    2}(n_4 a^i_4+n_5 a^i_5)(l_4 a^i_4+l_5 a^i_5)}
& &\hspace*{-5mm}\bar{\Theta}{u-2(n_4a^i_4+n_5a^i_5)
\atopwithdelims()v+2(l_4
  a^i_4+l_5 a^i_5)}(\taubar)\, ,\nonumber
\end{eqnarray}
where the integers $l_{4,5}$ correspond to $m_{4,5}$ of eq.~(\ref{mom22})
after
Poisson resummation. Performing a Poisson resummation back to $m_{4,5}$,
one finds that the $T^2$ momenta
$p_L,p_R$ for non-vanishing $a^i_{4,5}$ have 
the same form as in eq.~(\ref{mom22}) with the replacement:
\be
m_{4,5} \rightarrow m_{4,5}-\sum_{i=1}^{16}a_{4,5}^i
\left( k_i+{\alpha+\epsilon\over 4}\right) +
{1\over 2}\sum_{i=1}^{16}a_{4,5}^i(n_4a_4^i+n_5a_5^i)\ ,
\label{m'}
\en
where we defined $\alpha=2u-\epsilon=\pm 1$.
{}Furthermore, the partition functions ${\bar F}_P^{(s)}$ become:
\be
{\bar F}_P^{(\epsilon,\epsilon')} = \sum_{\alpha=\pm}
\alpha^{1+\epsilon'}\zeta_{\epsilon,\epsilon'}
{8\over\etabar^{18}\tetabar^2{\epsilon\atopwithdelims()\epsilon'}}
e^{-i\pi\taubar\sum_{i=1}^{16}
(k_i+{\alpha+\epsilon\over 4}-n_4a_4^i-n_5a_5^i)^2}\ ,
\nonumber
\label{FW}
\en
where $(\epsilon,\epsilon')=(1,0)$, $(0,1)$ or $(0,0)$ and the
coefficients
$\zeta_{\epsilon,\epsilon'}$ are given by:
\be
\zeta_{1,0}=-\cos \left({\pi\over 2}\sum_{i=1}^{16}k_i\right)\qquad
\zeta_{0,1}=-{1-(-)^{\sum_{i=1}^{16}k_i}\over 2}\qquad
\zeta_{0,0}=\sin \left({\pi\over 2}\sum_{i=1}^{16}k_i\right)\ .
\label{delta}
\en
As a result, the momenta $P$ of the $\Gamma_s^{(2,18)}$ lattices are 
characterized by
$4+16$ integers $m_{4,5}$, $n_{4,5}$ and $k_i$ associated to the $T^2$
coordinates and $U(16)$ right-moving fermions, respectively, as well as by
the
parameter $\alpha=\pm 1$ related to $u=0,1,\pm 1/2$ labelling the four 
conjugacy classes of $SU(16)$ level-one.

Since under heterotic -- type I duality the $T$-modulus is mapped into
$S'$,
in order to compare the above result with the type I expression
(\ref{KI1}), one
has to take the limit $\mbox{Im}T\to\infty$ \cite{ABFPT}. 
{}From eq.~(\ref{KH}) and the
expression of the $T^2$ momenta (\ref{mom22}), it follows that only
the Kaluza-Klein momenta corresponding to $n_4=n_5=0$ survive in this
limit.
Moreover, by making the change of variable 
\be
\tau_2={t\over 4}\mbox{Im}T\ ,
\label{t}
\en
one can
easily show that the integration domain becomes the strip 
$t\ge 0$, $-1/2\le\tau_1\le 1/2$, up to exponentially small corrections in 
$\mbox{Im}T$. We now perform the $\tau_1$ integration, which takes the
form:
\be
\int_{-1/2}^{1/2}d\tau_1 e^{-2i\pi\tau_1\left\{ \sum_{i=1}^{16}
(k_i+{\alpha+\epsilon\over 4})^2/2-{3+\epsilon\over 4}+{1+\epsilon\over
2}N\right\} }\, ,
\label{tau1}
\en
where $N$ is a non-negative integer coming from the oscillator expansion
of the
partition function (\ref{FW}). In the untwisted sector $\epsilon=1$ the
frequencies are integers and the intercept is at $-1$, while in the
twisted
sector $\epsilon=0$ the frequencies are half-integers and the intercept is
at
$-3/4$. An inspection of the exponent in eq.~(\ref{tau1})
using the expressions (\ref{delta}) shows that the coefficient of
$2i\pi\tau_1$
is always an integer. In fact, this follows immediately in the sector
$(1,0)$,
where $\epsilon=1$ and $\sum_i k_i$ is even, while in the remaining two
sectors where $\epsilon=0$ and $\sum_i k_i$ is odd it is half-integer.
However, since the sum of these sectors is proportional to 
$(1-\alpha (-)^{N+(1-\sum_i k_i)/2})$, one can show that only integer
values 
survive. As a result, the $\tau_1$ integration (\ref{tau1}) implies that
the
exponent vanishes:
\be
{1\over 2}\sum_{i=1}^{16}
\left( k_i+{\alpha+\epsilon\over 4}\right)^2-{3+\epsilon\over 4}+
{1+\epsilon\over 2}N=0\ .
\label{eq}
\en

Equation (\ref{eq}) in the $(1,0)$ sector can be satisfied only for
$\alpha=-1$;
the solutions are $N=1$ and $k_i=0$ or $N=0$ and $\sum_i k_i^2=2$.
In the sum of the other two sectors, it can be satisfied only for $N=0$;
the solutions in this case are $k_i=(0,\cdots ,0,\alpha,0,\cdots ,0)$.
Using eqs.~(\ref{KH}), (\ref{mom22}) and (\ref{m'}-\ref{delta}), we find:
\be
K^{H(1)}_{U\ubar}\rightarrow -{1\over (U-\ubar)^2}{1\over
8\pi^2\mbox{Im}T}
\int_0^\infty{dt\over t^2}\partial_t \left\{
16\sum_{\Gamma_2}\right. \, -
\!\!\!\!\!\!\sum_{{    {i<j} \atop {  {\varepsilon_1,\varepsilon_2=\pm 1}
\atop
{ {\varepsilon_1a^i+\varepsilon_2a^j+\Gamma_2} }}}}
\!\!\!\!\!\!\!\!\varepsilon_1\varepsilon_2 \; - \,8 \!\!\!\!\!\!\!\!\!
\sum_{ {i\atop{{\varepsilon=\pm 1}\atop{\varepsilon {\mbox{\small ( }}
\!\!a^i-\sum_j {a^j\over
4}{\mbox{\small ) }}\!\!\!+\Gamma_2}}}}
\!\!\!\!\!\!\!\!\! \left.
\phantom{\sum_{\Gamma_2}}\!\!\!\!\!\!\!\!\right\} 
t \ e^{-\pi t|p|^2/2}\ ,
\label{KHl}
\en
where $p$'s are given in eq.~(\ref{mom}). By introducing again the index
$I=i,\ibar$ defined in section 2 and using the 
traceless condition of $SU(16)$ generators $\sum_ja^j=0$,
it is easy to show that the third term inside
the bracket of eq.~(\ref{KHl}) can be written as $-8\sum_{a^I+\Gamma_2}$,
while the first two become
${1\over2}\sum^{I\ne J}_{a^I+a^J+\Gamma_2}$. Comparing with the
type I expression (\ref{KI1}), we conclude that the $(1,0)$ sector
reproduces
the contribution of the 99-open strings on the annulus and M{\"o}bius
strip, 
while the twisted sector $(0,1)+(0,0)$ reproduces the 95-contribution.

A similar analysis can be done for the threshold corrections to the
$SU(16)$ gauge
couplings $\Delta$, which are given by:
\begin{eqnarray}
\Delta^H &=& -{1\over 8}\sum_s\sum_{P\in\Gamma^{(2,18)}_s}
\int_{\cal F}{d^2\tau\over\tau_2}{\bar F}_P^{(\epsilon,\epsilon')}
\left\{ \left[ \sum_{i=1}^{16}\left( k_i+{\alpha+\epsilon\over
      4}-n_4a_4^i-n_5a_5^i\right) q_i\right] ^2
-{1\over 2\pi\tau_2} \right\} \nonumber\\ & &\qquad\qquad\times
e^{i\pi\tau|p_L|^2}e^{-i\pi\taubar|p_R|^2}\ ,
\label{DH}
\end{eqnarray}
in the notation of eqs.~(\ref{KH}), (\ref{mom22}) and (\ref{m'}-\ref{delta}).
Taking the limit $\mbox{Im}T\to\infty$ as before, one finds
\be
\Delta^H(T,U,A_i) \to c_0 \, \Im T+ \Delta^I(U,A_i) +4\pi^2
K^{I(1)}
\label{DHlim}
\en
up to exponentially
suppressed corrections. The last term, which originates from the
$1/2\pi\tau_2$ ``universal'' part of the threshold corrections
(\ref{DH}), is proportional to $1/\Im T$ and reproduces the type I
K{\"a}hler metric, eq.~(\ref{KI1}),  upon the replacement $T\rightarrow
S'$.  

The $T$-independent term $\Delta^I$ comes from the ``group-dependent'' 
part of threshold corrections (\ref{DH}) and is given by
\begin{eqnarray}
\Delta^I &=& {1\over 8} \int_0^{'\infty}{dt\over t}\left\{ 
\phantom{\sum_{\Gamma_2}} \right.
\!\!\!\!\!\!\!\!\!
\sum_{{    {i<j} \atop {  {\varepsilon_1,\varepsilon_2=\pm 1} \atop
{ {\varepsilon_1a^i+\varepsilon_2a^j+\Gamma_2} }}}}\!\!\!\!\!
\varepsilon_1\varepsilon_2
(\varepsilon_1 q_i +\varepsilon_2 q_j)^2 + 16\!\!\!\!
\sum_{ {i\atop{{\varepsilon=\pm 1}\atop{\varepsilon {\mbox{\small ( }}
\!\!a^i-\sum_{j=1}^{16} {a^j\over 4}{\mbox{\small )
}}+\Gamma_2}}}}\!\!\!\!
\left(q_i-\sum_{j=1}^{16}{q_j\over 4}\right)^2
\!\!\!\!\!\!\!\!\! \left. \phantom{\sum_{\Gamma_2}}\right\} e^{-\pi
  t|p|^2/2}\nonumber\\ 
&=& -{1\over 8}\int_0^{'\infty}{dt\over t} \left\{ 
\sum_{a^I+a^J+\Gamma_2}
\!\!\!\!\!s_{IJ}(q_I+q_J)^2\;
+\sum_{2a^I+\Gamma_2}(2q_I)^2 -\; 16\sum_{a^I+\Gamma_2}q_I^2
\right\} e^{-\pi t|p|^2/2}\ ,
\label{DHl}
\end{eqnarray}
where in the second line we used again the index $I=i,\ibar$ (such that
$q_{\ibar}=-q_i$) and the traceless condition of the $SU(16)$ generators
$\sum_jq_j=0$. 
The prime in the integral indicates that the apparent quadratic
divergence in the ultraviolet limit $t \to 0$ has been
subtracted. Equation (\ref{DHl}) then coincides with the threshold
corrections to the corresponding type I model \cite{BF}. There, the
ultraviolet divergence is removed by a regularization prescription,
which amounts to introducing a uniform cutoff in the transverse closed
string channel as dictated by the tadpole cancellation. Note that the
integrals (\ref{DH}) and (\ref{DHl}) are infrared finite due to
non-vanishing
Wilson lines.

The first term of eq.~(\ref{DHlim}) is linear in $\Im T$ with a
constant coefficient $c_0$ that controls the quadratic ultraviolet
divergence appearing in the limit $\Im T\to \infty$. In fact, $\Im T$
acts as a regulator in the heterotic integral (\ref{DH}) after the
change of variable (\ref{t}). In order to derive eq.~(\ref{DHlim}),
one goes back to the Poisson-resummed expression (\ref{modif}) with
$n_4=n_5=0$. Then, the term linear in $\Im T$ arises from the
$l_4=l_5=0$ contribution with 
\be
c_0=-{1\over 8}\int_{\cal F} {d^2\tau\over \tau_2^2} \sum_s \sum_{P\in
  \Gamma^{(16)}_s}\bar{F}^{(s)}_P \left\{ \left[
    \sum_{i=1}^{16}\left( k_i+{\alpha+\epsilon \over 4}\right) q_i
  \right]^2-{1\over 2\pi\tau_2} \right\} \, .
\label{c0}
\en
This term is mapped under duality to the part of the type I tree-level
gauge coupling proportional to $\Im S'$, cf.\
eqs.~(\ref{prei}-\ref{pretest})
for prepotentials. 
In the type I model under consideration, such a
term is absent, which implies that the integral (\ref{c0}) should
vanish. This is indeed the case as will be shown in the next section
when considering similar contributions to the gravitational couplings.

\section{Higher derivative F-terms}

We now consider the class of higher derivative F--terms of the form
\be
I_g=\F_gW^{2g}\ ,
\en
for integer $g\ge 0$ and $W$ being the Weyl superfield
\be
W_{\mu\nu}=F_{\mu\nu} - R_{\mu\nu\lambda\rho}\theta^1
\sigma_{\lambda\rho}\theta^2+ \cdots \ ,
\en
which is anti-self-dual in the Lorentz indices. 
$F_{\mu\nu}$ and $R_{\mu\nu\lambda\rho}$ are the (anti-self-dual)
graviphoton field strength and Riemann tensor, respectively. 
The couplings $\F_g$ are
holomorphic sections of degree $2-2g$ of the vector moduli, up to a
holomorphic anomaly for $g \geq 1$ \cite{BCOV}. 
They generalize the well-known prepotential $\F\equiv \F_0$, so that $\F_1$
determines the gravitational $R^2$ couplings etc.
On the type II side, $\F_g$ is determined entirely from genus $g$, while
on
the heterotic and type I, these couplings arise already at one loop (with
additional tree-level contibutions for $\F_0$ and $\F_1$). 

On the heterotic side, the torus
amplitude involving two gravitons and $(2g-2)$ graviphotons was used
in ref.~\cite{AGNT} to extract the $\F^H_g$ functions. The result involves
a
universal $2g$-point bosonic correlator
$G^H_g(\tau,\taubar)$ derived from the generating function 
\be
G^H(\lambda,\tau,\taubar) = \sum_{g=1}^{\infty}
\lambda^{2g}G^H_g(\tau,\taubar) = \left( \frac{2\pi i
\lambda \bar{\eta}^3}{\bar{\theta}_1(\lambda,\taubar)} \right) ^2 e^{-\pi
{\lambda^2 \over \tau_2}}+1\, ,
\label{G}
\en
which is modular invariant under $\displaystyle \tau \rightarrow {a\tau+b
\over
c\tau+d}$, $\displaystyle \lambda \rightarrow \pm {\lambda \over c\tau
+d}$. 
Here, $\theta_1$ is the odd theta function. Then, the generating function
\be
\F^H(\lambda,T,U,A_i)=\sum_{g=1}^{\infty} g^2 \lambda^{2g}
\F^H_g(T,U,A_i) \, ,
\label{Fgene}
\en
is given by \cite{AGNT,MS}:
\be
\F ^H(\lambda,T,U,A_i)={\lambda^2 \over 32\pi^2}
\sum_s\sum_{P\in\Gamma^{(2,18)}_s} 
\int_{\cal F}{d^2\tau\over\tau_2}{\bar F}_P^{(s)}(\taubar )
e^{i\pi\tau|p_L|^2}e^{-i\pi\taubar|p_R|^2}{d^2\over
  d\tilde{\lambda}^2} G^H(\tilde{\lambda},\tau,\taubar) \ ,
\label{FH}
\en
where we used the notation of eqs.~(\ref{KH}), (\ref{mom22}),
(\ref{m'}-\ref{delta}), (\ref{G}) and $\displaystyle
\tilde{\lambda}={\bar{p}_L 
\tau_2 \lambda \over \sqrt{2\, \mbox{Im}T \mbox{Im}U}}$.

Now, following the steps of section 3, we can take the limit 
$\mbox{Im}T\rightarrow\infty$:
\begin{eqnarray} 
\F ^H \to & &\hspace{-5mm}c_1 
\lambda^2 \Im T-{\lambda^2 \over 32} 
\int_0^{'\infty} {dt\over t}  \left\{
8\sum_{\Gamma_2}e^{-\pi t|p|^2/2}  
{d^2 \over d\lambda^2}\left[ \left({\lambda \over \sin \pi\tilde{\lambda}}
\right)^2 \right. \left(
2-\sin^2\pi\tilde{\lambda}\right)\right] 
\label{FHl}\\
& &-\hspace{-4mm}\sum_{{    {i<j} \atop {  
{\varepsilon_1,\varepsilon_2=\pm 1} \atop
{ {\varepsilon_1a^i+\varepsilon_2a^j+\Gamma_2} }}}}
\!\!\!\!\!\!\!\!\varepsilon_1\varepsilon_2e^{-\pi t|p|^2/2}  
{d^2 \over d\lambda^2}\left({\lambda \over \sin \pi\tilde{\lambda}}
\right)^2 \; - \,8 \!\!\!\!\!\!\!\!\!\!\!\!\!\!
\sum_{ {i\atop{{\varepsilon=\pm 1}\atop{\varepsilon {\mbox{\small ( }}
\!\!a^i-\sum_j {a^j\over
4}{\mbox{\small ) }}\!\!\!+\Gamma_2}}}}
\!\!\!\!\!\!\!\!\!e^{-\pi t|p|^2/2}{d^2 \over d\lambda^2} \left. 
\left({\lambda \over \sin \pi\tilde{\lambda}}
\right)^2 \right\} , \nonumber
\end{eqnarray}
where $p$ is given in eq.~(\ref{mom}) and after the change of variables
(\ref{t}) $\tilde{\lambda}$ becomes $\displaystyle \tilde{\lambda}=
{\bar{p} t
\lambda \over 4\sqrt{2\mbox{Im}U}}$. The first two sums on the r.h.s.
arise from the untwisted $(1,0)$ sector, whereas the third one is
due to the twisted $(0,1)+(0,0)$ sectors. Equation (\ref{FHl}) can be rewritten
as:
\begin{eqnarray} 
\F ^H &\to&
c_1 \lambda^2 \Im T -{\lambda^2 \over 32\pi^2}\int_0^{'\infty}{dt\over t} 
\left\{\sum_{a^I+a^J+\Gamma_2}\!\!\!\!\!s_{IJ}{d^2 \over
  d\lambda^2}\left({\pi\lambda \over \sin \pi\tilde{\lambda}}
\right)^2\right. 
\label{FI}\\
& &+\hspace{-2mm}\sum_{2a^I+\Gamma_2}{d^2 \over
  d\lambda^2}\left({\pi\lambda \over \sin \pi\tilde{\lambda}}
\right)^2 
-32\pi^2 \sum_{\Gamma_2}\left. -\; 
16\sum_{a^I+\Gamma_2}{d^2 \over
  d\lambda^2}\left({\pi\lambda\over\sin\pi\tilde{\lambda}}\right)^2\;
\right\} 
  e^{-\pi t|p|^2/2}\ . \nonumber
\end{eqnarray}

The above manipulations in taking the limit
$\mbox{Im}T\rightarrow\infty$ are valid for all $\F^H_g$ with $g\ge 2$ for
which
the $t$-integration converges. For $\F^H_1$ the integral diverges in the
infrared limit $t\to+\infty$, while it has also an apparent quadratic
ultraviolet divergence as $t\to 0$, which has been subtracted as
indicated by the prime. The infrared
divergence is already present in the original expression (\ref{FH}) before 
taking the limit and reproduces the trace anomaly of the effective field 
theory. Note that in contrast to the case of threshold corrections to
gauge 
couplings, the infrared divergence in the gravitational couplings cannot 
be regulated by non-vanishing Wilson lines since there are still massless 
states that contribute to the trace anomaly. In fact the gravitational 
beta-function \cite{AGN} is $b_{\mbox{\scriptsize
grav}}=(23+n_H-n_V)/12=2/3$,
which is reproduced by the infrared divergence arising from the third
and first terms (with $a^I+a^J=0$) in the integrand (\ref{FI}). On the
other hand, the ultraviolet divergence is regulated by $\mbox{Im}T$ as
discussed in the previous section, 
and gives rise to a term linear in $\mbox{Im}T$, which is mapped under
duality 
to a tree-level term proportional to $\Im S'$. Following the same
steps as in the derivation of $c_0$ in eq.~(\ref{c0}), one finds
\be
c_1 =-{1\over 8} \int_\F {d^2\tau\over \tau_2^2} \sum_s
\bar{F}^{(s)}\left(  
{2i\over \pi} \partial_{\taubar} \ln \etabar -
  {1\over 2\pi\tau_2} \right)\, ,
\label{c1}
\en
where we used eq.~(\ref{FH}) and 
\be
\left. {d^2\over d\lambda^2} G^H
\right|_{\lambda=0}=-8i\pi\partial_{\taubar} \ln \etabar +{2\pi \over
  \tau_2}\equiv -\bar{G}_2
\en
from eq.~(\ref{G}).

The integral (\ref{c1}) can be evaluated using the method of
ref.~\cite{LSW}. After replacing ${\displaystyle {1\over \tau_2}} =
{i\over \pi} \partial_{\taubar} \bar{G}_2$, the integral becomes a
total derivative and one is left with an integration along the
boundary of the fundamental domain $\F$. We can then easily show that
for arbitrary integer power $n$,
\be
\int_{\F}{d^2\tau \over \tau_2^2} \bar{G}_2^n F(\qbar) = {1\over
  (n+1)\pi} \left[ \bar{G}^{n+1}_2 F(\qbar) \right]_{\qbar^0} \, ,
\label{integral}
\en
where the subscript $\qbar^0$ denotes the constant 
term in the $\qbar=e^{-2i\pi\taubar}$ expansion. Using the above
formula for $n=1$ and the expansions
\be
\bar{G}_2={2\pi^2\over 3}(1-24 \qbar +\cdots)\ , \quad \sum_s
\bar{F}^{(s)}={2\over \qbar}(1-240 \qbar+\cdots)
\label{exp}
\en
following from eqs.~(\ref{Fbar}), we obtain
\be 
c_1=4\pi \, .
\label{c1v}
\en

After combining the one-loop contribution (\ref{c1v}) with the tree-level
term,
we obtain (in our conventions):
\be
\F^H_1=4\pi (\Im S + \Im T) + f_1^{H} \, ,
\label{F1Hpr}
\en
where $f_1^{H}$ represents the contribution contained in the regularized
integral of eq.~(\ref{FI}). After the replacement $T\to S'$, the
first two terms reproduce the tree-level type I expression for $\F_1^{I}$,
cf.\ eqs.~(\ref{gravh}-\ref{gravtest}).
Note the symmetry of this result under exchange of $S$ and $S'$, which is a
consequence of $T$-duality in the type I theory (inverting the volume of
$K_3$)
but a non-perturbative symmetry on the heterotic side. Finally $f_1^{H}$,
as
well as all higher $\F_g^{H}$'s encoded in the integral (\ref{FI}), should
be
compared with the corresponding one-loop contribution in type I theory.

Before discussing the type I contribution, we can compute at this point
the
constant $c_0$ entering in the expression of threshold corrections
(\ref{DHlim})
and given by the integral (\ref{c0}). Using the expressions (\ref{c0}) and
(\ref{c1}), it is easy to check that the difference $c_1-c_0$ is given by
an
integral of the form (\ref{integral}) with $n=0$:
\begin{eqnarray}
c_1-c_0 \!\!&=&\!\!{1\over 8}\int_{\cal F} {d^2\tau\over\tau_2^2}
\sum_s \sum_{P\in\Gamma^{(16)}_s}\bar{F}^{(s)}_P \left\{ \left[
\sum_{i=1}^{16}\left( k_i+{\alpha+\epsilon \over 4}\right) q_i\right]^2
-{2i\over \pi} \partial_{\taubar} \ln \etabar
\right\}\\
\!\!&=&\!\!{1\over 8}\int_{\cal F} {d^2\tau\over\tau_2^2}( 164-{1\over
  6}j(\qbar) )\, ,\nonumber
\end{eqnarray}
where $j$ is the modular invariant function with a simple pole 
at infinity with residue 1, $j(\qbar)=1/\qbar+744+\dots$ 
We now use eq.~(\ref{integral}) and the expansions (\ref{Fbar}),
(\ref{exp}) to obtain $c_1-c_0=4\pi$ and from the result (\ref{c1v}) 
we deduce
\be
c_0=0\, ,
\en
as was already announced in the previous section.

Let us now turn to the type I calculation of the $\F_g^I$'s. They can be
determined by the one-loop type I amplitude involving two gravitons
and $(2g-2)$ graviphotons \cite{GJNS,Se}. Following the steps of
ref.~\cite{Se} and using the results of ref.~\cite{ABFPT}, we obtain the
following general expression for the generating function (\ref{Fgene}):
\be
\F^I(\lambda)={\lambda^2\over 64\pi^2}\sum_{\sigma=\A,\M,\K}
\sum_{s,\Gamma_2^{(s)}}\int_0^\infty
{dt\over t}{\cal I}^{(s)}_{\sigma}e^{-\pi t|p|^2/2} {d^2\over
d\tilde{\lambda}^2} G^I_\sigma (\tilde{\lambda},t)+\F^I_{\mbox{\scriptsize
torus}}\ ,
\label{FIc}
\en
where the first sum is extended over the three different open string
surfaces,
annulus ($\A$), M{\"o}bius strip ($\M$) and Klein bottle ($\K$); 
$s$ denotes the
various sectors of the theory (orbifold as well as open string boundary
conditions), and
${\cal I}^{(s)}_{\sigma}$ is  an index associated to $K_3$ which counts the open
string spinors and closed
string Ramond-Ramond bosons weighted with the fermion-parity operator
$(-)^{F_{\mbox{\tiny int}}}$. 
The generating partition functions $G^I_\sigma$ were computed in 
refs.~\cite{GJNS,Se} and they have no explicit $t$-dependence:
\begin{eqnarray}
G^I_{\A}&=&G^I_{\M}=\left(
{\pi\tilde{\lambda} \over \sin \pi\tilde{\lambda}} \right)^2\ ,\nonumber\\ 
G^I_{\K}&=&\left( {2\pi \tilde{\lambda} \over \tan\pi \tilde{\lambda} }
\right)^2 = \left( {2\pi\tilde{\lambda}\over \sin\pi
\tilde{\lambda} } \right)^2 - 4\tilde{\lambda}^2\pi^2 \, . 
\label{GIs}
\end{eqnarray}

It is easy to see that as in the heterotic case, the $t$-integration
converges
for all $\F^I_g$ with $g\ge 2$. Moreover, $\F^I_1$ has an infrared
divergence 
as $t\to\infty$, which
reproduces the same trace anomaly $b_{\mbox{\scriptsize grav}}=2/3$
(after taking into account the torus contribution 
$\F^I_{\mbox{\scriptsize torus}}$ given below). 
On the other hand, the apparent
ultraviolet divergence at $t=0$ is cancelled among the contributions of
different surfaces when the integral is appropriately regularized as
discussed in section 3.

The last term in eq.~(\ref{FIc}) stands for the contribution of the
world-sheet
torus, which is non-vanishing only for $\F^I_1$ due to the odd-odd and
even-even
spin structures. The odd-odd contribution is \cite{Se}:
\be
\F^I_{\mbox{\scriptsize torus}}{\bigg |}_{\mbox{\scriptsize odd-odd}}
={\lambda^2\over 16}
\int_{\F} {d^2\tau \over \tau_2} \sum_{s,\Gamma^{(2,2)}}
{\cal I}^{(s)}_\T e^{i\pi \tau |p^{\mbox{\tiny II}}_L|^2} 
e^{-i\pi \taubar |p^{\mbox{\tiny II}}_R|^2}\, ,
\label{torus}
\en
where ${\cal I}^{(s)}_\T={\rm Tr}^{(s)}_R(-)^{F_{\rm int}}$ is 
the Witten index in the sector $s$ and
$p^{\mbox{\tiny II}}_L$, $p^{\mbox{\tiny II}}_R$ are given by
eq.~(\ref{mom22}) with the replacement $T\rightarrow i\sqrt{G}$.
The $\tau$-integration can be performed explicitly with the result
\cite{dkl}:
\be
\F^I_{\mbox{\scriptsize torus}}{\bigg |}_{\mbox{\scriptsize odd-odd}}
={\lambda^2\over 16}{\cal I}_{\T}
\left[ \ln\left( \makebox{Im}U|\eta(U)|^4\right) 
+\ln\left( \sqrt{G}|\eta(i\sqrt{G})|^4\right)+c'\right] \ ,
\label{int}
\en
where ${\cal I}_{\T}=\sum_s{\cal I}^{(s)}_\T$ and
$c'$ is an infinite constant due to the infrared divergence.
The contribution of even-even spin structures yields a similar
expression with a relative minus sign between the two terms in the r.h.s.
We thus find
\be
\F^I_{\mbox{\scriptsize torus}}={\lambda^2\over 8}{\cal I}_{\T}
\ln \left( \makebox{Im}U|\eta(U)|^4\right) \ ,
\label{torusf}
\en
up to a moduli independent additive constant. This result is independent 
of $\sqrt{G}$ and can be reproduced  by considering the integral 
(\ref{torus}) in the limit $\sqrt{G}\to\infty$.
The complex integration variable $\tau$ is then replaced by the real $t$
and the momenta become those of eq.~(\ref{mom}):
\be
\F^I_{\mbox{\scriptsize torus}}={\lambda^2\over 8}\sum_{\Gamma_2}
{\cal I}_{\T}
\int_0^\infty {dt\over t}e^{-\pi t|p|^2/2}\ ,
\label{torusfI}
\en
where the ultraviolet divergence at $t=0$ is regularized according to the 
type I prescription.

The second term in the r.h.s. of eq.~(\ref{int}) depends, in the context 
of type IIB, on the K{\"a}hler modulus of $T^2$ with its real part
projected 
away by the type I reduction. As was shown in ref.~\cite{BCOV} in general, 
this term does not couple to the gravitational
$R^2$ couplings, which depend only on the complex structure moduli. 
On the contrary, in type IIA the situation is reversed because the
even-even
contribution comes with a minus sign, and it
is the second term that survives. Moreover, under
duality these corrections are mapped into gravitational
instantons on the heterotic side \cite{HM2}. It is interesting to note 
that if such
corrections survive for some quantities in the context of type I strings, 
since under duality
$\sqrt{G}\to(\makebox{Im}T\makebox{Im}S/V_{K_3})^{1/2}$
(with $V_{K_3}$ the volume of $K_3$),
they are mapped on the heterotic side into stringy-like non-perturbative
effects
of the type $e^{-1/g}$.

Using the above results in eqs.~(\ref{FIc}), (\ref{GIs}) and
(\ref{torusfI}), 
we can write the following general expression for the generating 
function
\begin{eqnarray}
\F^I(\lambda)= & & {\lambda^2\over 64\pi^2}\sum_{s,\Gamma_2^{(s)}}
\int_0^\infty {dt\over t}({\cal I}^{(s)}_{\A}+{\cal I}^{(s)}_{\M}+4{\cal
I}^{(s)}_{\K})
e^{-\pi t|p|^2/2}{d^2\over d\lambda^2}
\left( {\lambda \pi\over\sin\pi\tilde{\lambda}} \right)^2\nonumber\\
& & +{\lambda^2\over 8}\int_0^\infty {dt\over t}
\sum_{s,\Gamma_2}
({\cal I}^{(s)}_\T-{\cal I}^{(s)}_\K)e^{-\pi t|p|^2/2}\ .
\label{FIs}
\end{eqnarray}
The sum $({\cal I}^{(s)}_{\A}+{\cal I}^{(s)}_{\M}+4{\cal I}^{(s)}_{\K})/4$ is
reduced to a sum over $N=2$ BPS hypermultiplets minus vector
multiplets (which come only from massless states in six dimensions),
hence for $g\ge 2$, $\F^I_g$'s 
are given  by expressions that are very similar to the 
prepotential \cite{ABFPT}.\footnote{See also the recent preprint quoted in
ref.~\cite{MS2}.} However, $\F^I_1$ has a different form
due to the presence of the second term in eq.~(\ref{FIs}).
In fact, the difference $({\cal I}^{(s)}_\T-{\cal I}^{(s)}_{\K})$ receives
contributions only from the (untwisted) $N=4$ sector of the theory,
which are already present at the level of the
corresponding $N=4$ supersymmetric type IIB theory \cite{HM2}. 

Specializing to the model under consideration, we obtain the following
contributions for the various open string boundary conditions:
\begin{eqnarray}
\mbox{99-sector} &:& \quad {1\over 4}
\sum_{s,\Gamma_2^{(s)}}{\cal I}^{(s)}_{\A}=  
-{1\over 2}\sum_{a^I+a^J+\Gamma_2}s_{IJ}\quad ;\quad {1\over 4}
\sum_{s,\Gamma_2^{(s)}}{\cal I}^{(s)}_{\M}=
-{1\over 2}\sum_{2a^I+\Gamma_2}\nonumber\\
\mbox{95-sector} &:& \quad {1\over 4}
\sum_{s,\Gamma_2^{(s)}}{\cal I}^{(s)}_{\A}=8\sum_{a^I+\Gamma_2}
\quad ;\quad{\cal I}^{(s)}_{\M}=0\\
\mbox{55-sector} &:& \quad {1\over 4}
\sum_{s,\Gamma_2^{(s)}}({\cal I}^{(s)}_{\A}+{\cal I}^{(s)}_{\M})=
-16\sum_{\Gamma_2}\ ,\nonumber
\label{special}
\end{eqnarray}
where the sum over $s$ refers to the orbifold sectors. On the other hand, in
the closed string sector we have ${\cal I}_{\K}=16$ and ${\cal I}_\T=24$ 
(the Euler number of $K_3$). Using these results we find that the
generating function (\ref{FIs}) yields the heterotic result
for $f_1^{H}$
[defined  in eq.~(\ref{F1Hpr})], as well as for all higher $\F_g^H$'s given 
in eq.~(\ref{FI}). The first two terms in the integral of the r.h.s.\
of eq.~(\ref{FI}) now correspond
to the  annulus and M{\"o}bius strip contribution in the 99-sector, the last
term to the contribution of the 95-sector and the third term to the
contribution of the torus.

\vspace{1cm}

\noindent{\bf Acknowledgements} 

We are grateful to C. Bachas and M. Serone for very useful conversations.

\vspace{1cm}

\end{document}